\documentclass[prc,aps,preprint,showpacs]{revtex4}
\usepackage{graphicx}
\begin{document}
\title{Properties of $^{12}$Be and 
$^{11}$Be in terms of single-particle motion in deformed potential} 

\author{ I. Hamamoto$^{1,2}$ and S. Shimoura$^{3}$ }

\affiliation{
$^{1}$ {\it Division of Mathematical Physics, Lund Institute of Technology 
at the University of Lund, Lund, Sweden} \\ 
$^{2}$ {\it The Niels Bohr Institute, Blegdamsvej 17, 
Copenhagen \O,
DK-2100, Denmark} \\ 
$^{3}$ {\it Center for Nuclear Study, University of Tokyo, Saitama 351-0198,
Japan} \\
}




\begin{abstract}
Inspired by the recent measurement of the 
B(E2;$0_{2}^{+} \rightarrow 2_{1}^+$)
and B(E0;$0_{2}^{+} \rightarrow 0_{1}^+$) values in $^{12}$Be, we give an
interpretation of available spectroscopic data on both $^{12}$Be and $^{11}$Be, 
using a simple model which contains the essential feature of these two nuclei;
the presence of weakly-bound neutron(s) in deformed potentials.
The agreement of the calculated results with observed data is surprisingly
good, including well-known strong E1 transitions in both nuclei.

\end{abstract}


\maketitle

\section{INTRODUCTION}
The study of the properties of nuclei far from the line of $\beta$ stability 
is currently one of the most active and challenging topics in 
nuclear structure.  Exotic ratios of Z/N for a given mass number together 
with the presence of weakly bound nucleons lead to the phenomena 
which are unexpected from our common sense about stable nuclei; 
among others, one finds the change of the shell structure and magic numbers.
Observed properties of low-lying states in
$^{11}_{4}$Be$_{7}$ and $^{12}_{4}$Be$_{8}$ have contributed to the finding 
that N=8 is 
no longer a magic number in such unstable nuclei.

Though there have been already many elaborate attempts to describe 
the structure of Be isotopes, 
in the present paper we give an interpretation of available 
spectroscopic data  
on $^{12}$Be and $^{11}$Be keeping a model as simple as possible, 
while taking duly into account the essential feature of these nuclei;  
the presence of weakly-bound neutron(s) in finite deformed potentials. 
A simplest version of Bohr-Mottelson model (see Chaps. 4 and 5 of Ref.
\cite{BM75}) is applied to the nuclei $^{11}$Be and $^{12}$Be, using 
weakly-bound one-particle wave-functions estimated properly with deformed
Woods-Saxon potentials instead of harmonic-oscillator potentials.
The present work is prompted by the recent lifetime measurements of the
0$_{2}^{+}$ state in $^{12}$Be \cite{SS06}. 

The deformation of Be isotopes has been studied using various
theoretical models; 
for example, $^{11}$Be is studied 
using models with deformed Woods-Saxon potentials 
\cite{EBS95, AM01},  
$^{11}$Be based on the Nilsson-Strutinsky method \cite{IR81}, and 
neutron-rich Be isotopes 
using the deformed Hartree-Fock calculation with Skyrme
interactions \cite{LH96}.  
The models used in Refs. \cite{EBS95, AM01}, in which the idea of the
particle-rotor model is used, may be closest to our present model. 
However, since in very light nuclei such as Be isotopes the nuclear shape 
may be considerably changed just by adding one neutron, 
we would avoid to use the observed properties of the core nucleus $^{10}$Be, 
in contrast to the work of Refs. \cite{EBS95, AM01}. 
Since only low-energy states are discussed in the present work, we
assume that four protons and six neutrons are treated as being inactive
occupying the lowest possible Nilsson levels for the same deformation as that
for the extra neutron(s).  
Thus, the inactive core 
is certainly different from the actual $^{10}$Be nucleus.

Moreover, the rotational perturbation of the intrinsic nuclear structure is
neglected since only those states close to the band head are studied in the
present work.  Thus, we try to 
see how the observed properties of low-lying levels of 
$^{11}$Be and $^{12}$Be can be understood in terms of weakly-bound neutron(s) 
in deformed potentials.
Deformed nuclear halos are studied in Ref. \cite{MNA97} using a finite
square-well potential without spin-orbit term, while in Ref. \cite{IH04} 
they are investigated with more realistic potentials: deformed Woods-Saxon
potentials with a spin-orbit term. 

In Sec. II some aspects of the model and formulas are summarized.
In Sec.III numerical results are presented and discussed in comparison with
experimental data, while 
in Sec. IV conclusions are given.

\section{MODEL AND FORMULAS}
It is known \cite{BM75} that in medium-heavy deformed nuclei the analysis of
low-energy spectroscopic properties in terms of one-particle motions in a
deformed potential works 
impressively well, especially if the pair-correlation in the form of
BCS quasiparticles is included. 
This is because the major part of the long-range two-body interaction 
is already taken into account in the deformed mean-field.   
In very light nuclei such as Be isotopes the many-body pair correlation 
that originates from a short-range interaction can be negligible.  
Therefore, in the present work we try to describe the spectroscopic properties
of low-lying states of $^{11}$Be and $^{12}$Be in terms of two Nilsson levels
occupied by the seventh neutron and the seventh
and eighth neutrons, respectively, in deformed Woods-Saxon potentials.
The pairing interaction between two weakly-bound neutrons in $^{12}$Be will be
duly taken into account.
This simple description may be justified, since there are no nearby Nilsson
levels in the region of deformation, where the [220,1/2] and [101,1/2] levels
are almost degenerate. 

One-particle levels with quantum numbers K$^{\pi}$ in a $Y_{20}$-deformed 
potential (Nilsson levels) are denoted by the asymptotic quantum numbers, 
[$N, n_z, \Lambda, \Omega$], where $\Omega = K$ and $\pi = (-1)^{N}$.  
The asymptotic quantum numbers express, in a good approximation, the wave
functions of one-particle levels well bound in sufficiently deformed
potentials.  When a one-particle level becomes weakly bound, the wave function 
can be very different from that indicated by the asymptotic quantum numbers, 
even when the one-particle potential is well deformed \cite{IH04}.
Nevertheless, just for convenience sake, 
in the present work 
we use the asymptotic quantum numbers to denote respective Nilsson levels, 
also in the case that those levels become weakly bound.

In the leading order the rotational energy spectrum is written as 
\begin{equation}
E_{rot} = \frac{\hbar^2}{2 \Im} \left( I(I+1) + a (-1)^{I+1/2} 
(I + \frac{1}{2}) \, \delta(K,1/2) \right)
\label{eq:erot}
\end {equation}
where $\Im$ expresses the moment of inertia while the decoupling parameter is
denoted by $a$.  The expression (\ref{eq:erot}), in which the rotational
perturbation of the intrinsic structure is neglected, is expected 
to work well for
low-lying rotational states close to the band head.

The formulas for B(E$\lambda$) to be used in the present work are 
(see Eqs. (4-91) and (4-92) of Ref. \cite{BM75})
\begin{eqnarray}
B(E \lambda ; K_1 I_1 \rightarrow K_2 I_2) & = & 
\{ C(I_1 \lambda I_2 ; K_1, K_2-K_1, K_2) \langle K_2 \mid E(\lambda, \mu 
= K_2-K_1) \mid K_1 \rangle \nonumber \\ 
& + & (-1)^{I_1+K_1} C(I_1 \lambda I_2 ; -K_1, K_1+K_2,
K_2) \langle K_2 \mid E(\lambda, \mu = K_1+K_2) \mid \tilde{K_1} \rangle 
\} ^2 \nonumber \\
&& \mbox{                 for   } K_1 \neq 0 \mbox{   and   } K_2\neq 0  
\label{eq:belam}
\end{eqnarray}
and
\begin{eqnarray}
B(E \lambda ; K_1 = 0, I_1 \rightarrow K_2 I_2) & = & 
\{ C(I_1 \lambda I_2 ; 0
K_2 K_2) \langle K_2 \mid E(\lambda, \mu = K_2) \mid K_1 = 0 \rangle \} ^2
\nonumber \\
& & \left\{ \begin{array}{lll}
2 & \qquad \mbox{for} & \qquad K_2 \neq 0 \\[0.1cm]
1 & \qquad \mbox{for} & \qquad K_2 = 0 
\end{array}
\right.
\label{eq:belam0}
\end{eqnarray}
where intrinsic matrix-elements are expressed by 
$\langle K_2 \mid E(\lambda, \mu) \mid K_1 \rangle$ .

From two different pairs of one-particle orbits ($\nu_1 \tilde{\nu_1}$) 
and ($\nu_2 \tilde{\nu_2}$) in deformed potentials, where $\tilde{\nu}$ 
expresses the time-reversed orbit of $\nu$, one can form two orthogonal
two-particle configurations 
\begin{eqnarray}
| K^{\pi} = 0_1^{+} \rangle & = & (a^2 + b^2)^{-1/2} (a|\nu_1 \tilde{\nu_1} 
\rangle + 
b| \nu_2 \tilde{\nu_2} \rangle ) \nonumber \\
| K^{\pi} = 0_2^{+} \rangle & = & (a^2 + b^2)^{-1/2} (-b|\nu_1 \tilde{\nu_1} 
\rangle + 
a| \nu_2 \tilde{\nu_2} \rangle )
\label{eq:ab}
\end{eqnarray}
This mixture of the two $|\nu \tilde{\nu} \rangle$ configurations 
can be interpreted as the result of pairing interaction. 
The quadrupole matrix element connecting these two states in the intrinsic
system is written as
\begin{equation}
\langle K^{\pi} = 0_2^{+} | \sum_{k} (r^2 Y_{20})_{k} | K^{\pi} = 0_1^{+} 
\rangle = \frac{2ab}{a^2 + b^2} 
( \langle \nu_2 | r^2 Y_{20} | \nu_2 \rangle - \langle \nu_1 | r^2Y_{20} | 
\nu_1 \rangle )
\end{equation}
Similarly, the monopole matrix element is 
\begin{equation}
\langle K^{\pi} = 0_2^{+} | \sum_{k} (r^2)_{k} | K^{\pi} = 0_1^{+} 
\rangle = \frac{2ab}{a^2 + b^2} 
( \langle \nu_2 | r^2 | \nu_2 \rangle - \langle \nu_1 | r^2 | 
\nu_1 \rangle )
\end{equation}
The unique feature of these transition matrix-elements 
is illuminated on p.552-553 of Ref. \cite{BM75}, in
connection with pair correlation and $\beta$ vibration.

In order to obtain the wave functions of one-particle levels in deformed
Woods-Saxon potentials, the coupled-channel equations for a given one-particle
level are solved \cite{IH04} in coordinate space with correct asymptotic
conditions for $r \rightarrow \infty$, without confining the system in a finite
box.

\section{NUMERICAL RESULTS}
\subsection{Energy and deformation parameter}
In $^{11}$Be 
known as a one-neutron halo nucleus the presence of an extremely 
strong E1 transition, $B(E1;1/2^{-} \rightarrow 1/2^{+})$ = $(0.115 \pm 0.01)$ 
e$^{2}$fm$^{2}$, is well known for years \cite{DJM83}.  
There are only two known bound states in $^{11}$Be: the 1/2$^+$ ground state 
and the 1/2$^-$ state at Ex=0.32 MeV. 
Due to the very small neutron separation 
energy $S(n)$=0.504 MeV, there is no hope to observe 
enhanced E2 transitions to establish possible rotational spectra based
on the ground state.  However, 
at least three resonant levels, 
5/2$^{+}$ at Ex = 1.78 MeV, 3/2$^{-}$ at Ex = 2.69 MeV and 3/2$^{+}$ 
at Ex = 3.41 MeV, have been established by recent experiments 
\cite{NF04,SDP06}.
If we regard the two one-particle resonant levels, 
5/2$^{+}$ at Ex = 1.78 MeV and 3/2$^{+}$ at Ex = 3.41 MeV, 
as members of the rotational band based on the ground state with 
$I^{\pi} = K^{\pi}$ = 1/2$^+$, 
we obtain the decoupling parameter $a$=1.82 and the rotational constant 
$\hbar^2 / 2 \Im$=0.403 MeV.  
Then, using the expression in Eq. (\ref{eq:erot}), 
the member of the ground-state rotational band in $^{11}$Be next lowest 
to 3/2$^+$ is the 9/2$^{+}$
state, which is expected around Ex=6.7 MeV in the absence of rotational
perturbation.  The centrifugal 
barrier for the 9/2$^{+}$ level is quite high and, thus, 
the possible resonance will be relatively sharp in spite of its higher energy. 

Assuming 
that the moment of inertia $\Im$ is proportional to $A^{5/3}$, 
$\hbar^2 / 2 \Im$=0.349 MeV is obtained 
for $^{12}$Be.  Then, we estimate the excitation
energy of the 2$_{1}^{+}$ state in $^{12}$Be to be 
\begin{eqnarray*}
E(2_{1}^{+}) = 6(0.349) = 2.09  \mbox{MeV,}  
\end{eqnarray*}
which indeed agrees with the observed value 2.11 MeV.
This agreement may indicate  
that the two nuclei, $^{12}$Be and $^{11}$Be, have similar deformations.

In $^{11}$Be 
we assume that the 1/2$^{+}$ and 1/2$^{-}$ levels are expressed by
the seventh neutron occupying 
one-neutron levels, [220,1/2] and [101,1/2], respectively, 
for a given strong prolate-deformation.   Using the standard 
spin-orbit strength \cite{BM67}, radius parameter $r_{0}$=1.25 fm, and a
diffuseness $a$=0.65 fm (potential [a]) or $a$=1.00 fm (potential [b]), 
the deformation $\beta$ and the depth of the Woods-Saxon 
potential $V_{WS}$ are determined so as to produce 
$\varepsilon$([220,1/2])=$-$0.5 MeV and 
$\varepsilon$([101,1/2])=$-$0.2 MeV.  
We obtain
\begin{eqnarray*}
\beta = 0.73, \qquad V_{WS} = -38.6 \: \mbox{MeV} \qquad & \mbox{for
potential [a]}   \\
\beta = 0.82, \qquad V_{WS} = -33.5 \: \mbox{MeV} \qquad & \mbox{for
potential [b]} 
\end{eqnarray*}
The $\beta$ value determined for $^{11}$Be in the literature depends on 
the used model and parameterization, and varies 
from 0.7 to 1.1 ; for example, see 
Ref. \cite{AM01}.
The diffuseness $a$=0.65 fm is a standard value for medium-heavy stable nuclei,
while $a$=1.00 fm is an attempt  
to simulate the potential without a flat bottom that may be commonly 
obtained in Hartree-Fock calculations for very light nuclei.  
In the present work the deformed Woods-Saxon potential 
is needed only to obtain the wave function 
of the seventh neutron, though the total 
neutron potential in the presence of a halo neutron may not be
properly described by a Woods-Saxon potential.
In Table \ref{tab:table1} 
we show the probabilities of $\ell_{j}$ components in the wave
functions of the [220,1/2] and [101,1/2] levels estimated for the potentials 
[a] and [b].  It is noted that the radial wave-functions of $\ell_{j}$
components of those deformed weakly-bound levels can be considerably different
from those of $n \ell_{j}$ eigenfunctions of spherical Woods-Saxon potentials
\cite{IH04,IH05}.  Therefore, great care has to be taken, when the 
spectroscopic factors are extracted in the analysis of the data such as the one
made in Refs. \cite{NF04,SDP06,AN00}.
In this sense, a caution is needed in 
the comparison of the $s_{1/2}$ probability, 0.62, 
in the [220,1/2] level calculated for the potential [b] with the spectroscopic
factor obtained in Ref. \cite{NF04}, 0.72$\pm$0.04 .

Using the [220,1/2] neutron wave-function for the potential [b] with 
$g_{R}$=Z/A and $g^{eff}_{s}$=0.9$g^{free}_{s}$, we obtain $-$1.72 $\mu_{N}$ 
for the magnetic
moment of the ground state of $^{11}$Be, which is in good agreement with the
observed value, $\mu$($^{11}$Be) = $-$1.6816(8) $\mu_{N}$ \cite{WG99}.  The
calculated value is not sensitive either to the used $g_{R}$ value of the
even-even core or the parameters of the Woods-Saxon potential. 
This insensitivity comes from the fact that the magnetic moment of an  
$s_{1/2}$ neutron is equal to 0.5$g^{eff}_{s}$, while in the asymptotic limit 
the magnetic moment of the I=1/2 state coming from the [220,1/2] neutron coupled
to the K=0 core is also equal to 0.5$g^{eff}_{s}$ independent of the core
$g_{R}$-value used.  In other words, in the present I=1/2 state the value of the
measured magnetic moment cannot be used to find 
whether or not the nucleus is deformed.

The spectroscopic properties of four known bound levels in $^{12}$Be, 
$0_{1}^{+}$, 
$2_{1}^{+}$ at Ex=2.11 MeV, $0_{2}^{+}$ at Ex=2.25 MeV and 
$1_{1}^{-}$ at Ex=2.70 MeV, have  
been experimentally studied for the last few years \cite{HI00,HI00a,SS03}.
In particular, the recent measurement of the lifetime of the $0_{2}^{+}$ level 
\cite{SS06} has pinned down  
the absolute magnitudes of $B(E2;0_{2}^{+} \rightarrow 2_{1}^{+})$ =
(7.0$\pm$0.6) e$^2$ fm$^2$ 
and $|\langle 0_{2}^+ |e \, r^2|0_{1}^+ \rangle |$ = (0.87$\pm$0.03) e fm$^2$, 
which are very valuable informations.
The relatively large values of 
$B(E2;0_{2}^{+} \rightarrow 2_{1}^{+})$ and 
$|\langle 0_{2}^+ |e \, r^2|0_{1}^+ \rangle |$
indicate that the intrinsic configurations of the $0_{1}^{+}$ and 
$0_{2}^{+}$ states are expressed by linear combinations 
of two two-neutron configurations, ($\nu_{1} \tilde{\nu_{1}}$) and 
($\nu_{2} \tilde{\nu_{2}}$), 
of which the intrinsic quadrupole moments are quite different. 
The analysis of 
inelastic proton scattering exciting the $2_{1}^{+}$ state in
inverse kinematics \cite{HI00} led to the deformation length, 
(2.00$\pm$0.23) fm, which suggested strong quadrupole deformation, $\beta
\sim 0.7 $.

Observed two 
0$^{+}$ levels in $^{12}$Be are assumed to be expressed by linear 
combinations (with equal amplitudes) of two two-neutron configurations, 
[220,1/2]$\widetilde{[220,1/2]}$ and [101,1/2]$\widetilde{[101,1/2]}$, 
for a given prolate deformation. 
The relation $|a| \approx |b|$ in the expression of Eq. (\ref{eq:ab})
is expected since the [202,1/2] and 
[101,1/2] levels in $^{11}$Be lie only 300 keV apart, following the
interpretation in the present model.
The deformation $\beta$ and the depth of the Woods-Saxon 
potential $V_{WS}$ are determined so as to produce both [220,1/2] and [101,1/2] 
levels at $-$1.1 MeV, which in the absence of the interaction 
between the two configurations is about a half of the binding energy of the
two neutrons.  Then, we obtain 
\begin{eqnarray*}
\beta = 0.66, \qquad V_{WS} = -39.0 \: \mbox{MeV} \qquad & \mbox{for
potential [a]}  \\
\beta = 0.72, \qquad V_{WS} = -35.2 \: \mbox{MeV} \qquad & \mbox{for
potential [b]} 
\end{eqnarray*}
The intrinsic configuration assumed here for the ground state of $^{12}$Be
is consistent with the fact that all four states, 
1/2$^+$, 1/2$^-$, 5/2$^+$ and 3/2$^-$, of $^{11}$Be are populated
in one-neutron removal reactions from $^{12}$Be with comparative magnitudes of
spectroscopic factors \cite{SDP06}.

\subsection{E1 transitions}
In medium-heavy stable nuclei low-energy E1 transitions
are so much hindered in both spherical and deformed nuclei 
that one could hardly expect to obtain a nuclear-structure
information from observed B(E1) values \cite{HHS93}.  The reason for the
hindrance is: (a) no appreciable amount of low-energy E1 strength 
due to the 
nuclear shell-structure; (b) the high-lying isovector giant dipole resonance 
absorbs the major part of possible E1 strength; (c) isoscalar dipole mode 
corresponds to the center of mass motion and thus does not contribute to E1 
excitations.
In contrast, in nuclei with weakly bound nucleons the origin of the hindrance 
(a) may be removed due to the change of the shell-structure 
or change in the one-particle wave
functions, while the hindrance (b) may be drastically reduced
due to the very weak coupling of the nucleons to the well-bound core.

In well-bound and well-deformed nuclei 
the E1 transition between the [220,1/2] and [101,1/2] levels, where
$| \Delta n_z |$=2 and $| \Delta \Lambda |$=1, 
is asymptotically forbidden.
Therefore, if the nucleus is well deformed, 
the observed strong E1 transition must come from
the properties of the levels related to being weakly bound.
It is pointed out that 
the $s_{1/2}$ component becomes
dominant in all $\Omega^{\pi}=1/2^{+}$ Nilsson levels as the binding energy 
of the orbits approaches zero, though it depends on respective Nilsson levels 
at which binding energy the dominance occurs \cite{IH04}.  
Since the major component of the [101,1/2] level is $p_{1/2}$, the $s_{1/2}$ 
dominance in the weakly-bound [220,1/2] level can produce 
a pretty strong E1 matrix-element.
In Table \ref{tab:table2} the calculated result of E1 properties is shown, 
where the neutron
E1 effective charge, $(Z/A)e$ = $(4/11)e$ = (0.364)e, 
is used neglecting the core
polarization effect 
(namely, the effect of shifting low-energy E1 strength to the  
high-lying isovector giant dipole resonance)
since halo neutrons may hardly polarize the well-bound core.  
The formula used to estimate the B(E1) value is  
\begin{eqnarray}
B(E1&;&I^{\pi}=K^{\pi}=1/2^{+} \rightarrow I^{\pi}=K^{\pi}=1/2^{-}) 
 \nonumber \\
& = &  (e_{eff}^{n}(E1))^2 
\{ C(\frac{1}{2},1,\frac{1}{2};\frac{1}{2},0,\frac{1}{2})
\langle [101,1/2] | rY_{10} | [220,1/2] \rangle \nonumber \\
 && \mbox{ } + (-1)^{1/2+1/2} C(\frac{1}{2},1,
\frac{1}{2}; \frac{-1}{2},1, \frac{1}{2} ) 
\langle [101,1/2] | rY_{11} | [\widetilde{220,1/2}] \rangle \}^2 
\end{eqnarray}
where $[\widetilde{220, 1/2}]$ expresses the time-reversed intrinsic
configuration of $[220, 1/2]$.
In the present case the contribution coming from 
the $Y_{11}$ term to the E1 matrix element 
(so-called 'signature-dependent term'), 
which originates from the $R$ symmetry
of the nuclear shape, is several times larger than 
that from the $Y_{10}$ term and
contributes coherently.
In both $Y_{10}$ and $Y_{11}$ matrix elements a smaller 
contribution from $d_{5/2} \rightarrow p_{3/2}$ 
contributes destructively to 
the major one from $s_{1/2} \rightarrow p_{1/2}$. 
The destructive structure is a remaining trace 
of the asymptotically forbidden E1
matrix-element. 
Observing that 
the B(E1) value obtained by using the potential [b] is in good agreement 
with the
observed value, (0.115$\pm$0.01) e$^{2}$ fm$^2$, 
in the following we show numerical
results obtained with the potential [b]. 

In the E1 transition, $0_{1}^{+} \rightarrow 1^{-}$, of $^{12}$Be
the $Y_{10}$ term contributes if the $I^{\pi}=1^{-}$ state has 
$K^{\pi}=0^{-}$, while the $Y_{11}$ term does if $K^{\pi}=1^{-}$.
In both $K^{\pi}=0^{-}$ and $K^{\pi}=1^{-}$ cases the band head 
will be $I^{\pi}=1^{-}$.  If $K^{\pi}=0^{-}$, the rotational
band has members with 
$I^{\pi} = 1^-, 3^-, 5^-, ...$, while the rotational band with 
$K^{\pi}=1^{-}$ has 
$I^{\pi} = 1^-, 2^-, 3^-, ...$   That means, 
if one observes the 2$^-$ rotational member among resonant
levels, $K^{\pi} = 1^-$ is assigned to the $I^{\pi} = 1^{-}$ level 
at Ex=2.70 MeV.  Since, 
to our knowledge, 
no resonant level that may be a candidate for the 
rotational member is yet found, in Table \ref{tab:table3} 
we show the calculated 
results of both $K^{\pi}=0^-$ and $1^-$ cases.  The intrinsic wave
functions used are 
\begin{eqnarray}
\Phi (K^{\pi} = 0_1^+) & = & \frac{1}{\sqrt{2}} ([220,1/2][\widetilde{220,1/2}] + 
[101,1/2][\widetilde{101,1/2}]) \nonumber \\
\Phi (K^{\pi} = 0^-) & = & \frac{1}{\sqrt{2}} ([101,1/2][\widetilde{220,1/2}] - 
[\widetilde{101,1/2}][220,1/2]) \nonumber \\
\Phi (K^{\pi} = 1^-) & = & [101,1/2][220,1/2]
\label{eq:wf01}
\end{eqnarray}
Then, the formulas used to estimate those B(E1) values are
\begin{eqnarray}
B(E1;I^{\pi} = K^{\pi} = 0_1^+ \rightarrow 
I^{\pi} = 1^-, K^{\pi} = 0^-) 
= 4 | \langle [101,1/2]| rY_{10} | [220,1/2] \rangle |^2 
( e_{eff}^{n} (E1))^2  \nonumber \\
\\
B(E1;I^{\pi} = K^{\pi} = 0_1^+ \rightarrow 
I^{\pi} = K^{\pi} = 1^-)         
=  4 | \langle [220,1/2]| rY_{11} | [\widetilde{101,1/2}] 
\rangle |^2 ( e_{eff}^{n} (E1) )^2 \nonumber \\
\end{eqnarray}
In Table \ref{tab:table3} 
$|e_{eff}^{n} (E1) | = (Z/A)$e = (4/12)e = (0.333)e is employed. 
Since the matrix-element of $rY_{11}$ is larger than that of $rY_{10}$, 
the calculated 
B(E1;$I^{\pi} = K^{\pi} = 0_1^+ \rightarrow I^{\pi} = K^{\pi} = 1^-$) value is
several times larger than 
the B(E1;$I^{\pi} = K^{\pi} = 0_1^+ \rightarrow 
I^{\pi} = 1^-, K^{\pi} = 0^-$) value.  Since $|e_{eff}^{n}|$ for E1 transitions 
for neutrons with the binding energy of $-$1.1 MeV 
may be somewhat smaller than 
(Z/A)e due to non-negligible core-polarization effect, 
the assignment of $K^{\pi} = 1^-$ to the $I^{\pi} = 1^-$ state at Ex=2.70 MeV 
is suggested 
from the comparison of the calculated and observed B(E1) values.

It is noted that in the present simple model, where $|a| = |b|$  
is
assumed in the expression (\ref{eq:ab}), both the 
B(E1; $I^{\pi} = K^{\pi} = 0^{+}_{2} \rightarrow I^{\pi} = 1^-, K^{\pi} = 0^-$)
and 
B(E1; $I^{\pi} = K^{\pi} = 0^{+}_{2} \rightarrow I^{\pi} = K^{\pi} = 1^-$)
values in $^{12}$Be vanish.  Thus, it is very interesting 
to measure the upper limit of
the E1 transition, 1$^-$ at 2.70 MeV $\leftrightarrow$ 0$^{+}_{2}$ at 2.25 MeV, 
though the small transition-energy makes the measurement very difficult.

\subsection{E2 transition in $^{12}$Be}
Now, we consider the E2 transition, 0$_{2}^{+}$ $\rightarrow$ 2$_{1}^{+}$, 
in $^{12}$Be.  
The 2$_{1}^+$ state at Ex = 2.11 MeV is interpreted as a rotational
member of the ground 0$_{1}^{+}$ state.
Using the intrinsic configuration
\begin{eqnarray}
\Phi (K^{\pi} = 0_2^+) & = & \frac{1}{\sqrt{2}} (-[220,1/2][\widetilde{220,1/2}] + 
[101,1/2][\widetilde{101,1/2}]) 
\label{eq:wf02}
\end{eqnarray}
and $\Phi (K^{\pi} = 0_1^+)$ in Eq. (\ref{eq:wf01}), we obtain
\begin {eqnarray}
B(E2 & ; & I^{\pi} = K^{\pi} = 0_{2}^{+} \rightarrow I^{\pi} = 2_{1}^{+}, 
K^{\pi} = 0_{1}^{+}) \nonumber \\
& = & 
(e_{eff}^{n}(E2))^2 \{ \langle [220,1/2] | r^2Y_{20} | [220,1/2] \rangle - 
\langle [101,1/2] | r^2Y_{20} | [101,1/2] \rangle \}^2  
\label{eq:e200}
\end{eqnarray}
Using the potential [b] for $^{12}$Be we obtain
\begin{eqnarray}
\langle [220,1/2] | r^2Y_{20} | [220,1/2] \rangle = 11.56 \quad \mbox{fm}^2 
\nonumber \\
\langle [101,1/2] | r^2Y_{20} | [101,1/2] \rangle = -0.32 \quad \mbox{fm}^2 
\label{eq:r2y2}
\end{eqnarray}
and, then,
\begin{eqnarray}
B(E2 ; I^{\pi} = K^{\pi} = 0_{2}^{+} \rightarrow I^{\pi} = 2_{1}^{+}, 
K^{\pi} = 0_{1}^{+}) 
& = & 141 \, (e_{eff}^{n}(E2))^2 \nonumber \\ 
& = & 5.6 \, \sim \, 12.7 \; \mbox{e$^2$fm$^4$}
\label{eq:be2}
\end{eqnarray} 
where $e_{eff}^{n}(E2)$ = (0.2 $\sim$ 0.3)e is used in the last line.
These values of $e_{eff}^{n}(E2)$ 
are obtained from the analysis of measured
quadrupole moments of $^{15}_{5}$B$_{10}$ and $^{17}_{5}$B$_{12}$ using the
shell model with the $0 \hbar \omega$ space \cite{SA01}.  
Both the large neutron
excess and the weak binding of neutrons make $e_{eff}^{n}(E2)$ smaller.
The estimated value (\ref{eq:be2}) is in agreement with 
the measured value, B(E2; $0_{2}^+ \rightarrow 2_{1}^+ $) = (7.0 $\pm$ 0.6) 
e$^2$fm$^4$.
It is noted that for the present large prolate deformation the matrix element 
$\langle [101,1/2] | r^2 Y_{20} | [101,1/2] \rangle$ is positive if the 
[101,1/2] level is deeply bound, in contrast to the negative value in
(\ref{eq:r2y2}).  On the other hand, 
for a more moderate prolate deformation the matrix
element is negative irrespective of one-particle energies. 
The dependence of the expectation values of $r^{2} Y_{20}$ in the [101,1/2] 
and [220,1/2] states on one-particle energies is exhibited in Fig. 1 
for two deformation values.
It is seen that in the region of 
$\varepsilon_{\Omega} < -1$ MeV the indication of
the divergence, 
$\langle [101,1/2] | r^{2} Y_{20} | [101,1/2] \rangle$ $\rightarrow$ $-\infty$ 
for $| \epsilon_{\Omega} | \rightarrow 0$, which comes from the behavior of 
$\ell = 1$ one-particle wave-functions 
for $| \varepsilon_{\Omega} | \rightarrow 0$ 
\cite{KR92}, has not yet appeared.  Thus, in the present case 
the calculated value of the 
expression (\ref{eq:e200}) is not sensitive to a small variation of
$\varepsilon_{\Omega}$.
The very small value of $\mid \langle [101,1/2] \mid r^2 Y_{20} \mid [101,1/2]
\rangle \mid$ for $\varepsilon_{\Omega} < -1$ MeV comes from the dominance of 
p$_{1/2}$ component in the [101,1/2] wave function, because the density
distribution of p$_{1/2}$ orbit is spherically symmetric.
On the other hand, for all values of $\varepsilon_{\Omega} < 0$ the major
contribution to $\langle [101,1/2] | r^{2} Y_{20} | [101,1/2] \rangle$ comes
from the $r^2 Y_{20}$ matrix element between the components $p_{1/2}$ and 
$p_{3/2}$ in the [101,1/2] wave function.
Note that the ratio of the probability of $p_{3/2}$ to that of $p_{1/2}$ 
in the [101,1/2] wave function has no strong variation 
for $| \varepsilon_{\Omega} | \rightarrow 0$.

In order to see the effect of weakly-bound neutrons, we estimate the
above B(E2) value using the wave functions in deformed harmonic-oscillator
potentials (see, for example, Ref. \cite{BP71}).
Noting that 
\begin{eqnarray}
\langle [220,1/2] | 2z^2 - x^2 - y^2 | [220,1/2] \rangle = 5 c_{z}^2 -
c_{\perp}^2 \\
\langle [101,1/2] | 2z^2 - x^2 - y^2 | [101,1/2] \rangle = c_{z}^2 -
2 c_{\perp}^2 
\label{eq:czp}
\end{eqnarray}
in the unit of $\hbar/m \omega_{0}(\epsilon)$, where 
\begin{eqnarray*}
c_{z}^2 = \frac{3}{3 - 2 \epsilon} \qquad \mbox{and} \qquad 
c_{\perp}^2 = \frac{3}{3 + \epsilon}
\end{eqnarray*}
and $\epsilon$ expresses the $Y_{20}$ deformation parameter of the 
oscillator potential, we obtain   
\begin{eqnarray}
\langle [220,1/2] | r^2Y_{20} | [220,1/2] \rangle  & - &
\langle [101,1/2] | r^2Y_{20} | [101,1/2] \rangle \nonumber \\
& = & \sqrt{\frac{5}{16 \pi}} \, (4 c_{z}^2 + c_{\perp}^2) \qquad 
\mbox{in unit of} \quad \hbar/m \omega_{0}(\epsilon) \nonumber \\
& = & 2.62 \quad \mbox{in unit of} \quad 
\hbar/m \omega_{0}(\epsilon) \nonumber \\
& = & 5.66 \quad \mbox{fm}^2 \qquad \mbox{for A=12}
\label{eq:tr2y2}
\end{eqnarray}
where $\epsilon$=0.7 is used and the volume conservation condition
\begin{equation}
\omega_{0}(\epsilon) = \omega_{00} \,(1-\frac{\epsilon^2}{3}-\frac{2}{27} \,
\epsilon^3)^{-1/3}
\end{equation}
is taken into account with $\hbar \omega_{00}$ = 41 A$^{-1/3}$ MeV.  
Using 
Eqs. (\ref{eq:r2y2}) and (\ref{eq:tr2y2}) it is seen 
that the weakly-bound character of two neutrons in $^{12}$Be makes the  
B(E2;$I^{\pi} = K^{\pi} = 0_{2}^{+} \rightarrow I^{\pi} = 2_{1}^{+}, 
K^{\pi} = 0_{1}^{+}$) value larger by 
(11.88/5.66)$^2$ = 4.41.
It is noted that the matrix element 
$\langle [101,1/2]|r^2 Y_{20}|[101,1/2] \rangle$ in the oscillator potential, 
which is given 
in (\ref{eq:czp}), changes the sign at $\epsilon$=0.6 and is
positive for $\epsilon$=0.7, in agreement with the sign change of the 
$r^2 Y_{20}$ matrix element of the deeply-bound [101,1/2] level shown 
in Fig. 1.

\subsection{E0 transition in $^{12}$Be}
For the E0 transition, 0$_{2}^{+}$ $\rightarrow$ 0$_{1}^{+}$, in $^{12}$Be  
we obtain 
\begin{eqnarray}
\langle [220, 1/2]| r^2 |[220,1/2] \rangle = 34.4  \: \mbox{fm}^2 \nonumber \\
\langle [101, 1/2]| r^2 |[101,1/2] \rangle = 23.0  \: \mbox{fm}^2
\end{eqnarray}
by using the wave functions calculated with the potential [b] for $^{12}$Be.
Then, using the intrinsic wave-functions, (\ref{eq:wf01}) and (\ref{eq:wf02}), 
we obtain  
\begin{equation}
| \langle 0_{1}^+ |r^2| 0_{2}^+ \rangle | = | \langle K^{\pi} = 0_{1}^{+} | 
r^2 | K^{\pi} = 0_{2}^{+} \rangle | = 11.4 \: \mbox{fm}^2 
\end{equation}
which is the neutron matrix-element.
The neutron effective charge for E0 transitions, $e_{eff}^n(E0)$,  
may sensitively depend on
both neutron excess and weak binding. 
The contribution to $e_{eff}^n(E0)$ coming 
from the part of the one-body operator obtained 
by subtracting the center of mass 
motion is $(Z/A^2)e$, which is equal to (0.028)e for $^{12}$Be. 
However, this may 
hardly give a reliable estimate of the actual $e_{eff}^n(E0)$ value.  
If the charge radius of $^{12}$Be relative to that of either $^{11}$Be 
or $^{10}$Be is ever measured, it will help to obtain an estimate of 
the $e_{eff}^n(E0)$ value to be used in the present case \cite{BM06}.  
Since no such measurement is presently available, writing the measured value as 
\begin{eqnarray*}
\langle 0_{2}^+ | e_{eff}^{n}(E0) \, r^2 | 0_{1}^+ \rangle = 0.87 \quad e \,
fm^{2}, 
\end{eqnarray*}
we obtain 
\begin{eqnarray}
e_{eff}^n (E0)/e = 0.87/11.4 = 0.076
\end{eqnarray}

\section{CONCLUSIONS}
We have presented the interpretation of available spectroscopic data on
$^{12}$Be and $^{11}$Be, using a model which is simple and contains the
essential feature of the presence of weakly-bound neutron(s) in deformed
potentials.  Calculated results are
in good agreement with available data.  We have intentionally avoided
to make a quantitative comparison of spectroscopic factors with "measured" 
ones, partly because the presently
available factors obtained from the analysis of data may, in our
opinion, contain 
an ambiguity in the one-particle radial wave-functions used and partly because
the inactive core nucleus $^{10}$Be in our present model is different from the
observed nucleus $^{10}$Be.
The features of weakly-bound neutrons appear especially in : (a) strong E1
transitions in both $^{11}$Be and $^{12}$Be due to the increased and
spatially-extended $s_{1/2}$ 
component in the [220,1/2] wave function; (b) The large 
B(E2;$I^{\pi}=K^{\pi}=0_{2}^{+} \rightarrow 
I^{\pi}=2_{1}^{+},K^{\pi}=0_{1}^{+}$) 
value, due to both the larger value of 
$\langle [220,1/2] | r^2 Y_{20} | [220,1/2] \rangle$ and the negative value of 
$\langle [101,1/2] | r^2 Y_{20} | [101,1/2] \rangle$ ; 
(c) The large $\langle [220,1/2] | r^2 | [220,1/2] \rangle$ matrix element that
contributes to the E0 transition, $0_{2}^{+} \rightarrow 0_{1}^{+}$.

In medium-heavy mass region one hardly finds a deformed nucleus, in which an
$\Omega^{\pi}$=1/2$^{+}$ level lies close to an $\Omega^{\pi}$=1/2$^-$ level 
around the ground state.  Therefore, the nucleus $^{11}$Be is the rare case, in
which $\Omega^{\pi}$=1/2$^+$ and 1/2$^-$ levels are almost degenerate and 
the importance of the presence of the signature-dependent term in B(E1) 
values is manifested.

One of the authors (I.H.) expresses her sincere thanks to 
professor Ben Mottelson for discussions.

\newpage

\begin{table}[htb]
\caption{\label{tab:table1} 
Calculated probabilities of $\ell_{j}$ components in the [220,1/2] and 
[101,1/2] wave functions of $^{11}$Be. 
Calculated radial wave-functions of the $\ell_{j}$ components are in general 
different from
those of the ($n \ell j$) wave functions which are the eigenfunctions of
spherical Woods-Saxon potentials.  
Two parameters of a given potential, deformation $\beta$ and the depth of
Woods-Saxon potentials, are adjusted so that the eigen energies of
the [220,1/2] and [101,1/2] orbits are $-$0.5 and $-$0.2 MeV, respectively. 
See the text for details.}
\vspace{0.3cm}
\begin{center}
\begin{tabular}{c|c|c|c|c|c|c|c|c|c|c} \hline \hline
  orbit & potential & \multicolumn{9}{c}{probability}   \\ 
  \cline{3-11} 
   &   & $s_{1/2}$  &  $d_{3/2}$  &  $d_{5/2}$  &  $g_{7/2}$  &  $g_{9/2}$  
   &  $p_{1/2}$  &  $p_{3/2}$  &  $f_{5/2}$  &  $f_{7/2}$ 
   \\ \hline 
 [220,1/2]  &  [a]  &  0.51  & 0.04 & 0.44 & 0.00 & 0.01  &  &  &  &   \\
  &  [b]  & 0.62   &   0.07  &  0.30 & 0.00 & 0.01 &  &   &   &   \\ \hline
 [101,1/2]  &  [a] &  &  &  &  & & 0.85  &  0.12 & 0.02 & 0.01  \\
  & [b] &  &  &  & & & 0.81  &  0.15 & 0.03 & 0.01  \\ \hline \hline  
\end{tabular}
\end{center}
\end{table}

\vspace{2cm}

\begin{table}[htb]
\caption{\label{tab:table2} 
Calculated results related to the E1 transition, ground 1/2$^+$ $\rightarrow$ 
1/2$^-$ at Ex=0.32 MeV, in $^{11}$Be.  For simplicity, in the table 
[220,1/2] and [101,1/2] 
are denoted by [$\nu_2$] and [$\nu_1$], respectively.  
The value of $e_{eff}^{n}(E1)$ = (Z/A)e
is used.  
The measured value in Ref. \cite{DJM83} is 
B(E1;$1/2^{+} \rightarrow 1/2^{-}$) = 
(0.115$\pm$0.01) e$^2$ fm$^2$.
See the text for details.}
\vspace{0.3cm}
\begin{center}
\begin{tabular}{c|c|c|c} \hline \hline
  potential & $\langle [\nu_1]|rY_{10}|[\nu_2] \rangle$ & 
  $\langle [\nu_1]|rY_{11}|[\nu_2]
  \rangle$ & 
  B(E1;$1/2^+ \rightarrow 1/2^-$)   \\  
   &   (fm) & (fm)&  (e$^2$ fm$^2$) \\ \hline 
 [a]  & -0.219 &  -0.615  &    (0.395) (e$_{eff}^{n}$(E1)/e)$^2$   \\
  &    &    &   $\Rightarrow$ 0.052   \\ \hline
 [b] & -0.355  &  -0.850 &  (0.809) (e$_{eff}^{n}$(E1)/e)$^2$  \\
  &    &    &   $\Rightarrow$ 0.107    \\ \hline \hline  
\end{tabular}
\end{center}
\end{table}

\newpage

\begin{table}[htb]
\caption{\label{tab:table3} 
Calculated results related to the E1 transition, ground 0$_{1}^+$ $\rightarrow$ 
1$^-$ at Ex=2.70 MeV, in $^{12}$Be.  The potential set [b] is used, while 
in the table [220,1/2] and [101,1/2] 
are denoted   by [$\nu_2$] and [$\nu_1$], respectively.  
The value of $e_{eff}^{n}(E1)$ = (Z/A)e
is used.  
The measured value in Ref. \cite{HI00a} is 
B(E1;$0_{1}^{+} \rightarrow 1^{-}$ = 
(0.051$\pm$0.013) e$^2$ fm$^2$.
See the text for details.}
\vspace{0.3cm}
\begin{center}
\begin{tabular}{c|c|c|c} \hline \hline
  $\langle [\nu_1]|rY_{10}|[\nu_2] \rangle$ & 
  $\langle [\nu_1]|rY_{11}|[\tilde{\nu_2}]
  \rangle$ &  
  B(E1;$0_{1}^+ \rightarrow I^{\pi}=1^-,K^{\pi}=0^-$) & 
  B(E1;$0_{1}^+ \rightarrow I^{\pi}=K^{\pi}=1^-$)   \\ 
  (fm) & (fm) & (e$^2$ fm$^2$)  &  (e$^2$ fm$^2$) \\ \hline
 -0.215  & -0.486 &  (0.184) (e$_{eff}^{n}$(E1)/e)$^2$  
 &    (0.944) (e$_{eff}^{n}$(E1)/e)$^2$   \\
  &    & $\Rightarrow$ 0.020   &   $\Rightarrow$ 0.105   \\ \hline \hline 
\end{tabular}
\end{center}
\end{table}

\newpage

\noindent
{\bf\large Figure captions}\\
\begin{description}
\item[{\rm Figure 1 :}]
The expectation value of $r^{2} Y_{20}$ in the [101,1/2] and [220,1/2] states
as a function of one-particle energy, $\varepsilon_{\Omega}$. 
For the [220,1/2] level the $s_{1/2}$, $d_{3/2}$, $d_{5/2}$, $g_{7/2}$ and 
$g_{9/2}$ components are included in solving the coupled-channel equations for
deformed one-particle levels,
while for the [101,1/2] level the $p_{1/2}$, $p_{3/2}$, $f_{5/2}$ and $f_{7/2}$ 
components are taken into account.  For comparison, two values of 
quadrupole deformation $\beta$ are considered.  Used parameters are $r_{0}$ =
1.25 fm, $a$ = 1.0 fm and $A = 12$, while the depth of Woods-Saxon potentials is
varied so that respective one-particle levels have $\varepsilon_{\Omega}$ 
as energy eigenvalues.

\end{description}

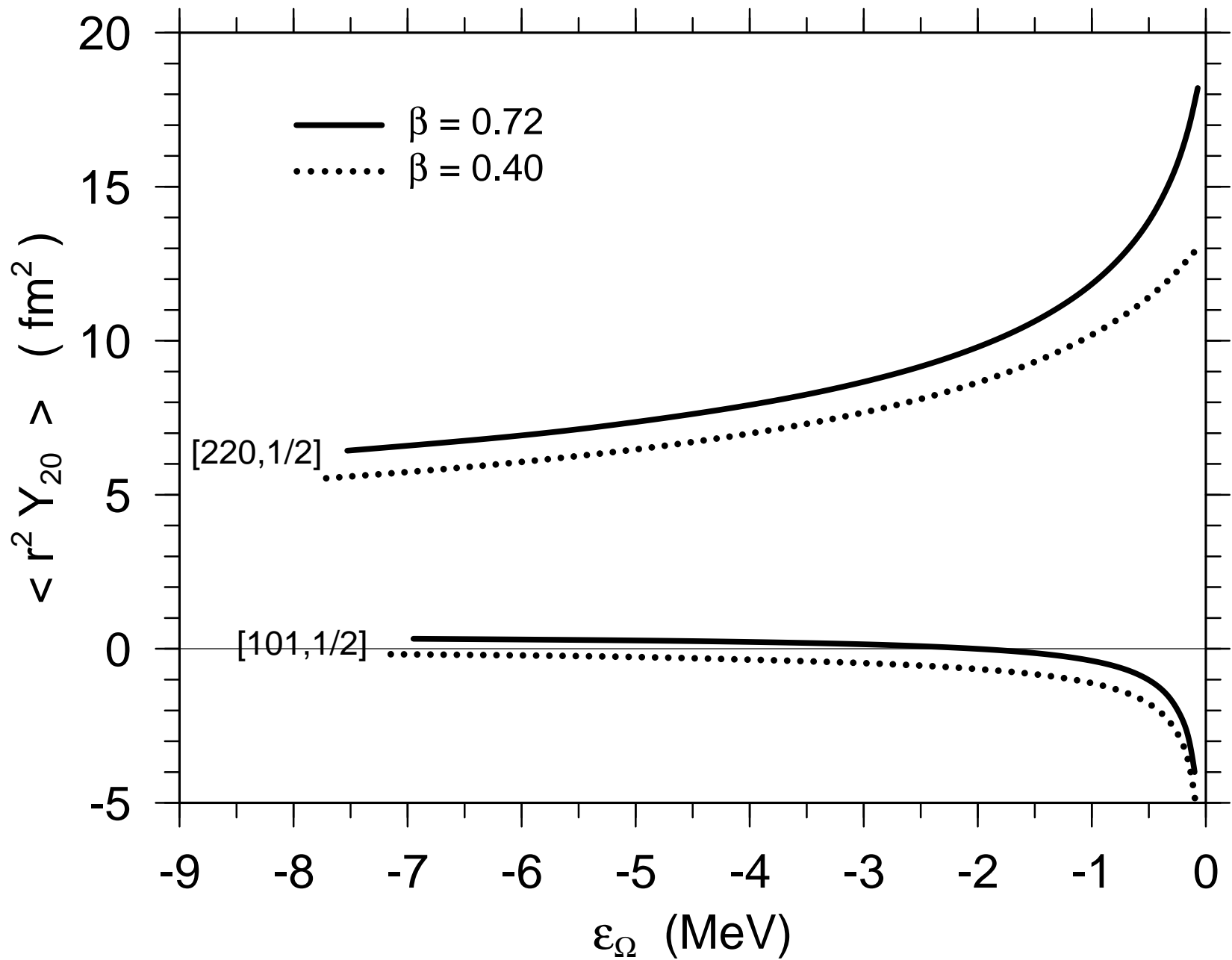
\end{document}